\documentstyle[prl,aps,psfig]{revtex}
\begin{document}
\draft
%%%%%%%%%%%%%%%%%%%%%
\twocolumn[
\title{%\vspace{-40pt}{\normalsize\null\hspace{5.5in}\sf DRAFT}\vspace{30pt}\\ 
Coexistence of, and Competition between,\\ Superconductivity and Charge-Stripe
Order in La$_{1.6-x}$Nd$_{0.4}$Sr$_x$CuO$_4$}
\author{J. M. Tranquada,$^1$ J. D. Axe,$^1$ N. Ichikawa,$^2$ 
A. R. Moodenbaugh,$^1$ Y. Nakamura,$^2$ and S. Uchida$^2$}
\address{$^1$Brookhaven National Laboratory, Upton, New York 11973}
\address{$^2$Department of Superconductivity, The University of Tokyo, Bunkyo-ku,
Tokyo 113, Japan}
\date{August 6, 1996}
\maketitle%
\widetext%
\advance\leftskip by 57pt \advance\rightskip by 57pt%
\begin{abstract}
Previously we have presented evidence for stripe order of holes and spins in
La$_{1.6-x}$Nd$_{0.4}$Sr$_x$CuO$_4$ with $x=0.12$.  Here we show, via
neutron diffraction measurements of magnetic scattering, that similar order 
occurs in crystals with $x=0.15$ and $0.20$.  Zero-field-cooled magnetization
measurements show that all 3 compositions are also superconducting, with the
superconducting transition temperature increasing as the low-temperature
staggered magnetization decreases.
\end{abstract}
]

\narrowtext

Neutron scattering studies
\cite{cheo91,maso92,thur92,yama95,aepp96,hayd96,yama96} of dynamical magnetic
correlations in superconducting La$_{2-x}$Sr$_x$CuO$_4$ have provided important
clues to the nature of electronic correlations within
the doped CuO$_2$ planes.  The low-energy magnetic scattering, which is
characterized by the two-dimensional antiferromagnetic wavevector 
${\bf Q}_{\rm AF}=(\frac12,\frac12)$ (measured in units $2\pi/a$) at low doping,
shifts to positions
$(\frac12\pm\epsilon,\frac12)$ and $(\frac12,\frac12\pm\epsilon)$, with
$\epsilon\approx x$ for $x>0.05$ \cite{yama96}.  In one common interpretation
\cite{bulu90,si93,litt93,tana94}, the incommensurate peaks are viewed as the
dynamical response of a spatially uniform electron liquid with a nearly-nested
Fermi surface.  From a rather different perspective, the
$Q$-dependent structure is taken as evidence for spatial inhomogeneity
associated with charge segregation \cite{low94,kive94,chay96} or
charge-density-wave correlations \cite{zaan89,schu89,poil89,cast95}.  Evidence
for the latter picture is provided by our recent discovery
\cite{tran95a,tran96b} of incommensurate charge and spin order in
La$_{1.6-x}$Nd$_{0.4}$Sr$_x$CuO$_4$ with
$x=0.12$; however, given the %lack of evidence for superconductivity down to a
%temperature of 5~K in the original crystal \cite{naka92}, 
claim \cite{buch94a} that bulk superconductivity is absent at this composition,
one might choose to
argue that these results are not directly relevant to the case of
superconducting samples.

To test the relationship between charge-stripe order and superconductivity, we
have now investigated two other Sr concentrations, $x=0.15$ and 0.20.  Our
neutron diffraction measurements on single-crystal samples reveal {\it elastic}
incommensurate magnetic peaks for both compositions, thus demonstrating the
presence of charge-stripe order.  Since the $x=0.20$ crystal was known to be
superconducting from previous work \cite{naka92}, we decided to check the
$x=0.12$ and 0.15 crystals for superconductivity as well.  To our surprise,
zero-field-cooled susceptibility measurements exhibit a bulk shielding signal for
all three compositions.  Since both the incommensurate peak splitting,
$\epsilon$, and the superconducting transition temperature vary with $x$, the
results strongly suggest a local coexistence of superconductivity and stripe
order.  The fact that $T_c$ decreases as the staggered magnetization increases
indicates that these two types of order compete with one another \cite{imry75}. 
Furthermore, since the variation of $\epsilon$ with $x$ in the Nd-doped
crystals is essentially identical to that obtained from recent inelastic
measurements \cite{yama96} on crystals of La$_{2-x}$Sr$_x$CuO$_4$, it seems 
inescapable that dynamical charge-stripe correlations are present in
the optimally doped material.

The crystals studied in this work were grown at the University of Tokyo using
the traveling-solvent floating-zone method.  The transport properties of the
$x=0.12$ and 0.20 compositions were reported several years ago \cite{naka92}; the
$x=0.15$ and further $x=0.12$ crystals were grown more recently.  The neutron
diffraction measurements on the $x=0.15$ and 0.20 crystals were performed on
triple-axis spectrometers at the High-Flux Beam Reactor, Brookhaven National
Laboratory, utilizing cryostats and spectrometer conditions similar to those
used in the previous work on $x=0.12$, which is described in detail elsewhere
\cite{tran95a,tran96b}.

\begin{figure}
\centerline{\psfig{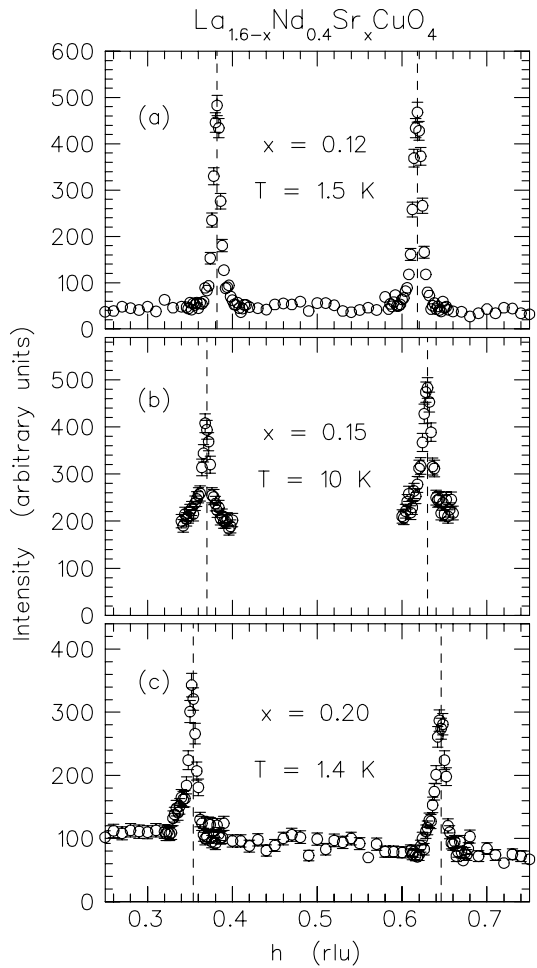}}
\bigskip
\caption{Scans along ${\bf Q}=(h,\frac12,0)$ through the magnetic peaks at
$h=\frac12\pm\epsilon$ measured on crystals of
La$_{1.6-x}$Nd$_{0.4}$Sr$_x$CuO$_4$ with (a) $x=0.12$, (b) $x=0.15$, and (c)
$x=0.20$.  Note that the measurements are not all at the same temperature.
\label{fg:1}}
\end{figure}

\begin{figure}
\centerline{\psfig{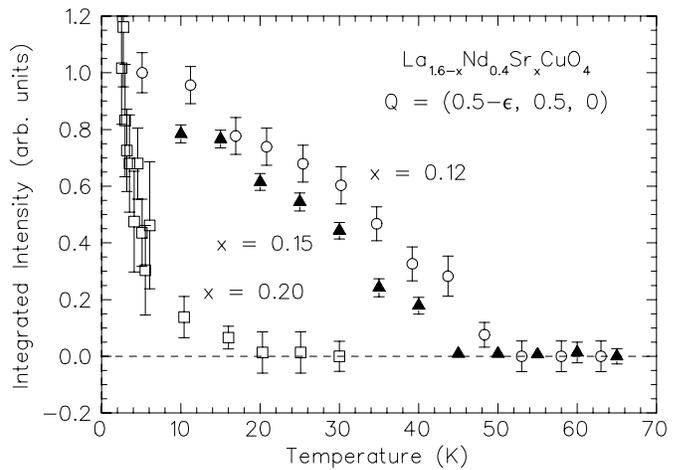}}
\bigskip
\caption{Temperature dependence of the incommensurate magnetic peak intensity
for crystals with $x=0.12$ (circles; Ref.~\protect\onlinecite{tran96b}), 0.15
(triangles), and 0.20 (squares).  Intensities are normalized for sample volume.
\label{fg:2}}
\end{figure}

Scans through the magnetic peaks at ${\bf Q}=(\frac12\pm\epsilon,\frac12,0)$
are shown in Fig.~\ref{fg:1}.  Sharp elastic peaks (with resolution-limited
widths in these coarse-resolution scans) are found for all three Sr
concentrations.  The peak splitting parameter,
$\epsilon$, is distinctly different in each sample, and clearly increases with
$x$.  The temperature dependences of the magnetic peak intensities (normalized
to sample volume) are presented in Fig.~\ref{fg:2}.  Both the ordering
temperature and the relative intensity (proportional to the square of the
staggered magnetization) decrease with
$x$.  The sharp upturn in intensity at low $T$ that is apparent for the $x=0.20$
sample is identical to that found previously for $x=0.12$
\cite{tran95a,tran96b}, and is due to ordering of the Nd moments via coupling to
the Cu ions.  The Nd ordering provides a useful amplification of the Cu order.

Unfortunately, there is no such incidental amplification of the charge-order
peaks, which were already quite weak for $x=0.12$.  An extremely weak signal
was detected at the expected position $(2+2\epsilon,0,0)$ for the $x=0.15$
crystal at 10~K, but it was not practical to determine its temperature
dependence.  No search for a charge-order peak in the $x=0.20$ sample was even
attempted due to the small size of the crystal ($\sim0.05$~cm$^3$) and to the
weakness of the magnetic signal.  Nevertheless, even without a direct
observation of charge order (or, rather, the corresponding lattice modulation to
which neutrons are sensitive), a modulation of the charge density is implied by
the incommensurate magnetic order.  The argument behind this assertion is as
follows.  The magnetic incommensurability indicates there there exists a
modulation of either the spin orientations (spiral order) or the spin density
\cite{zach96}; a combination of these two is also possible.  We have argued
elsewhere \cite{tran96b} that the secondary ordering of the Nd moments is
incompatible with perfect spiral order of the Cu spins within a plane; therefore,
there must be a spin-density-wave component to the order.  Symmetry allows a
spin modulation with wave vector {\bf q} to couple to a charge modulation at
2{\bf q}.  It follows that a charge-density modulation must be present; the
only real issues concern the magnitude of the modulation and the driving
mechanism.  In the case of $x=0.12$, the neutron diffraction data indicate that
the order is driven by the charge \cite{tran95a,tran96b,zach96}.  There is
nothing to suggest that the physics is any different in the $x=0.15$ and 0.20
crystals.

\begin{figure}
\centerline{\psfig{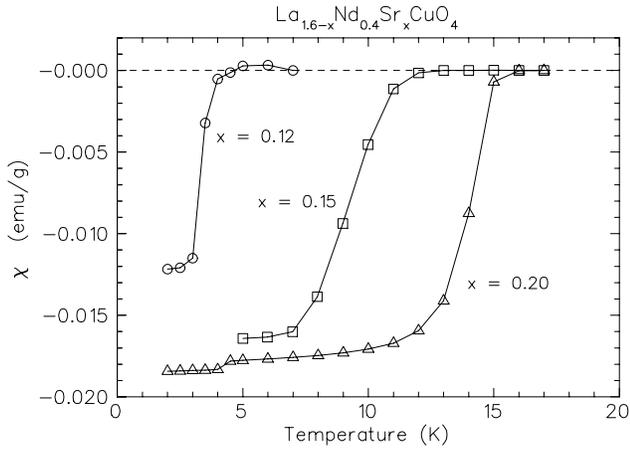}}
\bigskip
\caption{Bulk magnetic susceptibility measured after cooling in zero field, for
crystals with $x=0.12$, 0.15, and 0.20.  The kink at 4~K for $x=0.20$ is
attributed to hysteresis in the magnet. 
\label{fg:3}}
\end{figure}

To test for superconductivity in the crystals (or pieces thereof), the bulk
magnetic susceptibility was measured with a SQUID (superconducting
quantum-interference device) magnetometer, using a magnetic field in the
range of 1--5~G.  Attempts to measure the Meissner effect (by cooling in a
magnetic field) yielded a weak paramagnetic upturn at
$T_c$.  On the other hand, measurements performed after cooling in zero field
(see Fig.~\ref{fg:3}) give a shielding signal $>100$\%\ (without correction
for demagnetization).  Meissner-effect measurements on large samples are
notoriously difficult due to flux-pinning effects, whereas shielding 
measurements tend to be less problematic
\cite{naga93}.  We believe that the shielding results provide reliable evidence
of bulk superconductivity in the crystals.  The variation of $T_c$ with $x$
argues against associating the superconductivity with an impurity phase.

\begin{figure}
\centerline{\psfig{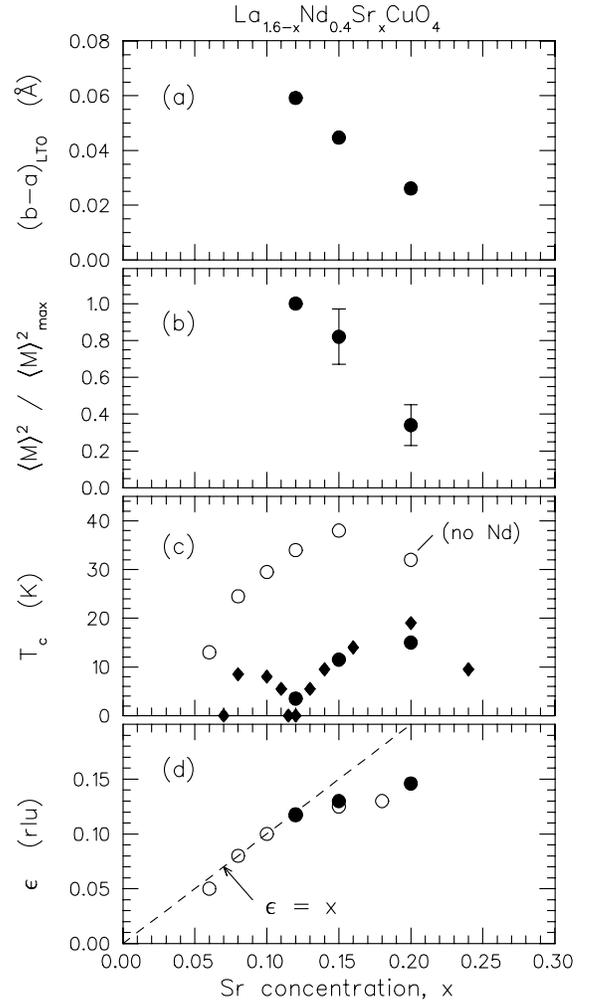}}
\bigskip
\caption{Comparison of results as a function of Sr concentration: (a)
difference between $a$ and $b$ lattice parameters in the LTO phase measured just
above the transition to the LTT phase, (b) square of the
low-temperature staggered magnetization, normalized to the $x=0.12$ result, (c)
superconducting transition temperature, and (d) incommensurate splitting,
$\epsilon$.  Filled symbols: La$_{1.6-x}$Nd$_{0.4}$Sr$_x$CuO$_4$; open symbols:
La$_{2-x}$Sr$_x$CuO$_4$ (Ref.~\protect\onlinecite{yama96}).  Circles:
single-crystal samples; diamonds: ceramic samples. 
\label{fg:4}}
\end{figure}

There have been disagreements in the literature \cite{craw91,buch94a} concerning
the existence of bulk superconductivity in La$_{1.6-x}$Nd$_{0.4}$Sr$_x$CuO$_4$
for $x\lesssim0.2$.  In particular, B\"uchner {\it et al.} \cite{buch94a} have
argued against bulk superconductivity on the basis of Meissner-effect and
specific-heat measurements.  We have already mentioned the difficulties with
Meissner measurements.  For the specific heat, it is observed in the cuprates
that the jump at $T_c$ rapidly becomes smeared as doping conditions deviate
from optimal \cite{lora89}; hence, it is not surprising if a superconducting
transition is not readily apparent in specific-heat measurements on a sample
with a severely depressed $T_c$.  As a check on the present single-crystal
results, a series of ceramic samples was prepared.  Shielding measurements
confirm the existence of bulk superconductivity, except at $x=0.07$, 0.115, and
0.12. The variation of $T_c$ with $x$ is compared with the single-crystal
results in Fig.~\ref{fg:4}(c).  The behaviors are reasonably consistent.  The
values of
$T_c$ for single crystals of La$_{2-x}$Sr$_x$CuO$_4$ studied recently by Yamada
{\it et al.} \cite{yama96} are also included (open circles).

The reduction of $T_c$ induced by the Nd substitution is correlated with a
modification of the low temperature tilt pattern of the CuO$_6$ octahedra
\cite{craw91}.  The Nd causes a change in the tilt
direction from [110], as in the low-temperature-orthorhombic (LTO) phase, to
[100], characteristic of the low-temperature-tetragonal (LTT) phase, with the
transition occuring at a temperature of roughly 70~K.  A coupling between
the tilt modulation and the charge-stripe correlations is possible only when the
tilts have a [100] orientation, parallel to the charge modulation.  One might
expect that the degree to which the charge modulations can be pinned would
depend on the amplitude of octahedral tilts.  B\"uchner {\it et al.}
\cite{buch94a} have shown that a useful measure of the tilt amplitude (or
actually its square) is the maximum difference between $a$ and $b$ lattice
parameters in the LTO phase.  The values of $b-a$ measured by neutron
diffraction on our Nd-doped crystals are shown as a function of Sr
concentration in Fig.~4(a); for comparison, the square of the staggered
magnetization (low-temperature magnetic peak intensity normalized relative to
the $x=0.12$ result) is presented in (b).  The strength of the magnetic order is
clearly correlated with the size of the tilt modulation, consistent with the
pinning argument.  The magnitude of
$T_c$ reduction is also correlated with the tilt modulation.  Superconductivity
and stripe order compete with each other, but also coexist.

Of course, the hole concentration also varies with $x$, and this is reflected
in the variation of the magnetic-peak-splitting parameter, $\epsilon$, as
indicated by the filled circles in Fig.~\ref{fg:4}(d).  The open circles are
the results of inelastic measurements on crystals of La$_{2-x}$Sr$_x$CuO$_4$ by
Yamada {\it et al.} \cite{yama96}.  The trends with $x$ are essentially
identical, implying that the nature of the instantaneous correlations in
the two systems is the same.  The recent observation \cite{hayd96} that
high-energy spin fluctuations in La$_{1.86}$Sr$_{0.14}$CuO$_4$ behave like damped
spin waves certainly seems consistent with the presence of stripe correlations.
The only signficant difference between the crystals with and without Nd appears
to be the degree of pinning of the stripe correlations.  Fluctuations of the
stripe correlations seem to be important for achieving a high $T_c$.

Theoretically, calculations by Vierti\"o and Rice \cite{vier94} indicate that
charged domain walls in a doped antiferromagnet will tend to melt due to
quantum fluctuations.  This result is quite consistent with experiment: no
static stripe order is observed in optimally-doped La$_{2-x}$Sr$_x$CuO$_4$. 
Quantum melting is inhibited only when a sufficiently strong perturbation, such
as that caused by Nd substitution, is applied.  The tendency of domain walls to
fluctuate has also been considered in other recent studies \cite{zaan96,eske96}.
Fluctuating stripe correlations seem a likely explanation for the quantum
critical behavior found in La$_{1.86}$Sr$_{0.14}$CuO$_4$ by Aeppli {\it et al.}
\cite{aepp96}.

The spatial modulation of spin and charge densities indicated by our results
could be driven either by a Fermi-surface-induced charge-density-wave (CDW)
instability or by frustrated phase separation.  CDW order is generally
stabilized by the opening of a gap about the Fermi energy.  Such a gap would
seem to be inconsistent with the low resistivity \cite{naka92} and
superconductivity in these samples; furthermore, optical measurements show no
evidence for a gap in charge excitations down to 4~meV \cite{taji96}.  On the
other hand, in the frustrated-phase-separation model\cite{kive94,kive96}, the
chemical potential should lie within a band of mid-gap states, consistent with
metallic behavior.  (An alternative strong-correlation model with charge stripes
is described in \cite{naya96}.)

To summarize, we have presented evidence that superconductivity and charge-stripe
order coexist in La$_{1.6-x}$Nd$_{0.4}$Sr$_x$CuO$_4$, although the order
parameters compete with one another.  For a given $x$, the spatial modulation
of the spin correlations is the same as in La$_{2-x}$Sr$_x$CuO$_4$
\cite{yama96}, which indicates that the instantaneous correlations are
essentially the same in the two systems.  There appears to be an intimate
connection between stripe correlations and superconductivity in these
materials.  It will be interesting to test the generality of these results in
other families of cuprate superconductors.

We gratefully acknowledge stimulating discussions with R. J. Birgeneau, V. J.
Emery, S. A. Kivelson, M. A. Kastner, G. Shirane, F. Wilczek, K. Yamada, and O.
Zachar.  We also thank K. Yamada for sharing his results prior to publication. 
Work at Brookhaven was carried out under Contract No.\ DE-AC02-76CH00016,
Division of Materials Sciences, U.S. Department of Energy.

%\bibliographystyle{prsty}
%\bibliography{lno,theory,optical}

\newpage

\newpage

\newpage

\newpage

\end{document}